# Data Science in Perspective

Rogério Rossi
Center for Mathematics, Computing and Cognition (CMCC)
Federal University of ABC (UFABC)
Santo André, São Paulo, Brazil
rogeriorossi8@gmail.com

*Abstract*—Data and Science has stood out in the generation of results, whether in the projects of the scientific domain or business domain. CERN Project, Scientific Institutes, companies like Walmart, Google, Apple, among others, need data to present their results and make predictions in the competitive data world. Data and Science are words that together culminated in a globally recognized term called Data Science. Data Science is in its initial phase, possibly being part of formal sciences and also being presented as part of applied sciences, capable of generating value and supporting decision making. Data Science considers science and, consequently, the scientific method to promote decision making through data intelligence. In many cases, the application of the method (or part of it) is considered in Data Science projects in scientific domain (social sciences, bioinformatics, geospatial projects) or business domain (finance, logistic, retail), among others. In this sense, this article addresses the perspectives of Data Science as a multidisciplinary area, considering science and the scientific method, and its formal structure which integrate Statistics, Computer Science, and Business Science, also taking into account Artificial Intelligence, emphasizing Machine Learning, among others. The article also deals with the perspective of applied Data Science, since Data Science is used for generating value through scientific and business projects. Data Science persona is also discussed in the article, concerning the education of Data Science professionals and its corresponding profiles, since its projection changes the field of data in the world.

*Keywords-Applied Data Science; Data Science; Data Science Persona; Formal Data Science*

## 1. INTRODUCTION

Data and Science can be considered integrated *per se*, since science depends on data in order to present its results. Data and Science words favor the presentation of a term that is currently globally recognized as Data Science. Data Science has transformed scientific, academic, and business environments in many aspects; by its structuring phase, by its interdisciplinarity, by the necessity of combining new technologies and massive data, and by the necessity of specialized professionals.

Many concepts referring to Data Science is proposed by [1], [2] and [3]. For [4] "Data Science might therefore imply a focus involving data and, by extension, statistics, or the systematic study of the organization, properties, and analysis of data and its role in inference, including our confidence in the inference".

Data Science can also be considered as a new formal knowledge area [2] that is in its initial phase of conceptualization and foundation, strongly integrating Statistics and Computer Science [5], also considering Artificial Intelligence, Machine Learning and Business Science. It is also possible to highlight Data Science as an applied discipline [6] which addresses the management and analysis of data in its various phases, whether for projects in the scientific or business domains.

Data Science is a new field, focused on the process and systems that enable the organization to extract knowledge or insight from data and translate it into action [7], which emphasizes Data Science as an applied area capable of generating results in organizations of different industries, categories or size.

Formal or applied, Data Science has to be well structured, both in the aspect of managing data (attribution of the Data Science Professional), as well as in the aspect of decision making based on achieved results (attribution of the decision makers).

Many questions are posed as more people seek Data Science knowledge: What are the foundations of Data Science? What are the activities and methods used in the Data Science area? And what are the activities of a Data Science Professional? What are the tools and methods that enable Data Science feasible? In this sense, students of undergraduate and graduate programs; professionals who are in the industry and who work with Data Intelligence; academics and scientists, who are involved in data analysis; business executives from different sectors and levels; they seek answers regarding Data Science, about its concepts, the interdisciplinarity that involves Data Science, the technologies and techniques, the tools and methods that are capable of making Data Science practicable.

In addition, two relevant questions can be highlighted: 1. How 'science' impacts formal Data Science? and, 2. How Data Science is applied in Scientific and Business Projects? These questions permeate this research, which using the exploratory method, address the relevance of 'science' for formal and applied Data Science, highlighting the use of 'science' and the scientific method for Data Science; either as a formal knowledge area, which is based on formal sciences (Mathematics, Statistics and Computer Theory), and as an applied area, to support decision making, based on Business Science, Data Management and (Big) Data Mining.

Science implies knowledge gained through systematically study [4]. Science is strictly linked to scientific method; science is built from scientific research and through scientific method and with scientific resources. Science takes place through the scientific method that can be associated with several areas that are guided by the scientific domain, whether in formal, natural, social or applied sciences.

Considering that science also implies aspects of curiosity and systematization, specific to a scientific researcher or a scientist, [8] considers that a Data Scientist, in addition to technical expertise, must be curious to distill problems into a clear set of hypotheses that can be testable and [9] states that Data Scientist is the sexiest job of the 21st century.

Based on this introduction that move toward Data Science, Scientific Method and Data Science projects in Scientific and Business domain, this article is structured as follows: section two discusses the integration of scientific method and Data Science; section three presents the aspects of Data Science as formal and applied area; section four deals with Data Science Professional, its education and its main activities; and, section five presents the final thoughts.

## 2. Data Science and the Scientific Method

This section introduces some applicable issues on the scientific method in order to observe that this method can be considered as part of Data Science activities, as an area that needs to accomplish its activities in an organized and systematic way.

Contemporary science considers specific procedures that are connected to the scientific method. The method assists the researcher's actions in order to determine the path and activities to be followed to generate scientific results that, in the end, can be validated or refuted.

This section does not address the issues regarding the scientific method and how it is used and applied nowadays. The main objective is simply the approach that a method exists to guide scientific activities and, consequently, the activities of a scientist, emphasizing that it can also be applied by a Data Science Professional.

Science implies organization and systematization to generate knowledge [4], thus, this statement implies a strong relationship with Data Science; to begin with because Data Science supports knowledge generation from data; and, in addition, because a systematic method can be applied in Data Science activities.

Accordingly [10], the scientific method is built around testable hypotheses. Models are tested and experiments confirm or falsify them. This statement summarizes how science has been treated for years. For [11] the scientific method attempts to remove from the subjective domain using hypotheses to be tested to validate models and eventually improving knowledge. These statements generate significant convergences to Data Science activities, even if the scientific method is used partially or, in a customized way, to manage Data Science activities.

Dodig-Crnkovic [12] presents a logical scheme used to produce scientific theories. These are the steps that compose the scientific method: 1. Pose the question in the context of existing knowledge; 2. Formulate a hypothesis; 3. Deduce consequences and make predictions; 4. Test the hypothesis; and 5. Provide a set of propositions to define a new phenomenon or a new theoretical concept. Considering this method, it is possible to verify that the first four steps provide important subsidies for Data Science, as well as for Data Science professionals, who can base their activities according to them.

The scientific method is associated with the determination of new theories or phenomena, it is carried out to offer results according to data analysis, since science requires data to determine its results. In the same way, it occurs for Data Science, which in the scope of business, is not bound to generate new theories or phenomena but, from data intelligence, support decision-making. Thus, data are fundamental in the scientific domain for the foundation of new theories and, in the business domain, to support decision making; strengthening the approach of using the scientific method for Data Science.

Boyd and Crawford [11] states that researchers interpret data and have linked the domain of research and the scientific method to Big Data, given that massive and varied datasets can provide better results from analyzes carried out by researchers from various scientific fields, be it natural sciences, social sciences, or logic and mathematics.

Following the analysis of the relationship between science, the scientific method and massive data, [10] considers that the conceptual approach to science that deals with - hypothesize, model and test - (Hypothetical-Deductive Scientific Method) has become obsolete, considering that statistical algorithms are able to find patterns in massive datasets where science could not find them.

Van der Aalst [13] also considers that new researchers and scientists have primarily used data analysis and interpretation and not models, possibly due to the amount of data available and the ease use of data analysis tools.

Agarwal and Dhar [14] point out that whereas the scientific process corresponds to a cycle of hypothesis generation, experimentation, hypothesis testing and inference, with several starting points, massive data (or Big Data) can be useful both in generation and in hypothesis testing.

Brodie [15] considers that the scientific method favors data analysis when it comes to a small number of variables associated with the research domain or the phenomenon to be evaluated. However, for Data Science, in some problems, an unlimited number of correlations between a large set of variables associated with the problem can be considered, thus carrying out the analysis due to the high computational capacity and the efficiency of the algorithms, not necessarily by the method.

EDISON Data Science Framework (EDSF) [16] also deals with the application of Scientific Method to Data Science. EDSF presents in the Data Science Competence Framework (CF-DS) [17] a Competence Group called 'Research Method and Project Management' which emphasizes as main objective "Create new understanding and capabilities by using the scientific method (hypothesis, test/artefact, evaluation) or similar engineering methods to discover new approaches to create new knowledge and achieve research or organizational goals". This Competence Group determines the use of the scientific method or similar engineering methods for Data Science, emphasizing the use of specific methods as part of Data Science activities.

In this sense, the questions about the application of scientific method to Data Science activities are relevant, they are also sometimes conflicting. [2] mentions an agenda for Data Science, considering that governments, industries, research and education institutions seek to promote Data Science as a new field of science.

## 3. DATA SCIENCE – FORMAL AND APPLIED

Data Science should become a formal area of knowledge as part of formal sciences such as Mathematics, Statistics and Computer Science and, on the other hand, Data Science tends also to present itself as an applied area, capable of generating value from the analysis of massive data through powerful computational algorithms.

In this way, this section mentions initial foundations regarding Data Science as formal science, presenting some specific issues that denote the structuring of the Data Science area. The section also discusses Data Science as an applied area that is useful for data intelligence in institutions, companies or corporations of any domain, category or size.

Considering Data Science as a formal area of knowledge, for [15] Data Science is in its infancy, [2] also mention that Data Science is in its initial phase, and [18] considers that Data Science is in an embryonic stage. For [5], Data Science is the child of Statistics and Computer Science, as well as for [19]. [20] adds that in its embryonic phase, Data Science must also consider the aspects of Business Sciences. These statements present some convergence and they also allow us to ponder that, in some way, that Data Science is not yet a structured and sedimented formal science.

Cleveland [19] presents a plan for technical work in the field of statistics, and mentions that given the substantial change in this field, including multidisciplinary research, this would imply a new field, called Data Science. [19] also presents a set of actions as a way to implement this plan, each of these actions with percentages of impact to the expansion plan of the technical areas of the statistics field evolving to what can be called Data Science, emphasizing the creation and the structure of a new formal area of knowledge nominated Data Science.

Cao [2] presents an evolutionary view of Data Science as an area of knowledge, considering a natural evolution of Data Science from Statistics and addressing some of the main (key) terms in Data Science, such as: Advanced Analytics, Data Analytics, Predictive Analytics, Deep Analytics. [2] adds that the evolution from Data Analysis to Data Science begins with the community of mathematicians and statisticians.

Van der Aalst [13] mentions three main ingredients for Data Science: 1) Infrastructure (Big Data, Distributed Systems, programming, etc.), 2) Analysis (Statistics, Machine Learning, etc.), and 3) Effect (business models, operations, etc.). The Analysis ingredient determines the formal part of Data Science and, in addition to these ingredients, [13] presents a vision of Data Science applied in several domains (medicine, logistics, social sciences, etc.) emphasizing Data Science as part of applied sciences.

In this sense, Data Science as an applied area must be able to generate value and results, whether in the scientific domain or business domain, and implies an organization of human and non-human resources, such as: methods, processes, tools, technologies and concepts from Statistics, Computer Science and Machine Learning as part of AI.

Data Science implies the capacity that the decision maker has on the data, thus requiring the science of data intelligence acquired based on data value chain that can be considered, but is not limited to these phases: Data Collection, Data Preparation, Data Analysis and Data Visualization. This general value chain for data lifecycle has favored the proposal of different methods, workflows, or pipelines to be used as part of Data Science activities.

Brodie [15] highlights the proposal of a specific method for Data Science activities - 'A generic Data Science Method'. For [15], the method was created based on the scientific method, as this is currently the method capable of supporting Data Science activities. The method considers eight steps that make Data Science activities feasible, highlighting steps that deal with the formulation of the problem, the formulation of hypotheses, the validation of the model until the conclusion and the validation of the results.

At a high conceptual level, it is possible to verify a correspondence considering the scientific method presented by [12] and the Generic Data Science Method presented by [15], which, in a formal perspective, links the concept of science, scientific method and Data Science.

Blei and Smyth [5] also proposes a set of activities favorable for managing Data Science activities. A possible pipeline can be derived from these activities: 1) understanding a problem domain; 2) deciding which data to acquire and how to process it; 3) exploring and visualizing the data; 4) selecting appropriate statistical models and computational methods, and 5) communicating the results of the analysis.

Matsudaira [21] emphasizes that in order to generate results, Data Science, as an applied area, needs a clear process to deliver better. [21] also mentions that a group of Data Scientists from a Data Science area proposes ideas, investigates hunches, and tests hypotheses, and that these are complex activities that imply difficulties in estimating the work to be performed and guaranteeing the results. [21] adds that the steps of an established process for Data Science activities cannot be based on existing process models, such as those applied to Software Engineering, such as the Agile Methods, like Scrum, because the research activities associated with Data Science do not correspond to the activities proposed by these recent methods. This is yet another example that points to the need for a real closer approximation of 'science' concept, scientific method and Data Science, since Data Science is able to generate value, in the scientific and business domain projects.

In closing, further studies and in-depth research are required to form, through science, the concepts and foundations of Data Science. Based on this, theories, concepts, and methods could offer better conditions to applied Data Science. Data Science, as a formal area, has its initial concepts and theories determined by Statistics and Computer Theory, however its strong interdisciplinarity also integrates other disciplines, such as Artificial Intelligence, with more emphasis on Machine Learning methods and models, Business Science, Data Mining, among others.

## 4. DATA SCIENCE PERSONA

For Oberski [22] Data Science is mostly about humans. The Data Science professional, commonly called Data Scientist, is the professional who holds the knowledge regarding Data Science activities, methods, and technologies to be applied in projects that are carried out in scientific or business domains, in order to generate value and results that support the organization's intelligence actions.

A diversity of professional profiles for Data Science can be verified in the Data Science Profession Profile (DSPP) [23], a component of EDSF framework [16], but possibly the most common is Data Scientist. Whether in the managerial, technical, or specialist areas, there are specific professional profiles for Data Science, such as: Data Science Manager, Data Stewards, Large Scale Database Designers, Big Data Facilities Operator, among others.

Smith [1] consider that Data Science include "the study of the capture of data, their analysis, metadata, fast retrieval, archiving, exchange, mining to find unexpected knowledge and data relationships, visualization in two- and three-dimensions including movement, and management". This specific definition presents a set of activities that can be considered for a Data Science Professional. For the specific concept of Data Scientist, [24] considers that it is an emerging job title to people who are able to tackle big data problems. [25] considers Data Scientist as a practitioner with business domain expertise as also analytical skills, and programming and systems expertise to manage scientific methods in the big data lifecycle to deliver value to science or industry.

Many questions are posed about Data Science professionals - What is their ideal education? What degree of knowledge is required to be a Data Scientist? What are the abilities and responsibilities of a Data Scientist? How to train people for this profession? What scientific domain should they have? These questions have gradually been answered by the academic community or by the industry, and have enabled the generation of professionals to handle and analyze data.

Many efforts can be noted to meet the goal of Data Science education, whether by the industry itself, by governments, by academia, by associations and educational institutions. [2], [7], [22], [26], [27] collaborate for a better understanding of the efforts to address issues related to Data Science and education.

Song and Zhu [24] and Brunner and Kim [28] are more specifically concerned with issues related to how to teach Data Science; [26] concerns with the formative structuring of knowledge and profiles for Data Science professionals presenting the EDISON Data Science Framework (EDSF) [16] through the EDISON Project [29].

As [24] in the US, Data Science programs are available in four categories: Bachelor, Master, Specialization, and Certificate Programs. For the authors, Bachelor's Programs are at the beginning and most well-known universities that were investigated are often programs at the graduate level.

Cao [2] is emphatic in proposing the relationship between data education and data innovation and economy. "Think with data", "Manage data", "Mine on data" - represents the requirements for industry and governments to recognize the value of data for decision making. [2] presents some possible activities for a Data Scientist: 1. Learn the business problem domain, 2. Understand data complexity, 3. Set up analytical processes, 4. Transform business problems into analytical tasks, 5. Mine relevant data, 6. Write coherent reports and presentations – among others.

Patil [8], emphasizes the possible nomenclatures adopted for Data Science professionals, such as "Business Analyst", "Data Analyst" and also mentions the "Research Scientist", which has been a nomenclature used by some companies, but considers that the ideal term found for these professionals is "Data Scientist", as they use data and science to create something new. [8] proposes some characteristics for the profile of a Data Science Professional: 1. Technical expertise - deep expertise in some specific discipline, 2. Curiosity - distill a problem into a very clear set of hypotheses that can be tested, 3. Storytelling - using data to tell a story, 4. Cleverness - look at a problem in creative ways.

In order to provide the results based on data and science, specific competences proposed according to the EDSF [16] can be verified, specifically according to the Data Science Competence Framework (CF-DS) [17], which refers to the fact that Data Scientist must use research methods and principles in developing data driven applications and implementing the whole lifecycle of data handling.

## 5. CONCLUSIONS

Data and Science are two words that together form the term Data Science, more than just two words or one term, Data Science generates value through data intelligence in scientific research institutes and business organizations from different domains, size or category.

Data Science can represent a connection of 'science', the scientific method and (big) data. This connection can be perceived by the term 'Data Science' *per se*, but in fact, 'science' can formalize and structure what is Data Science today or what it will be in the future. The foundations of Data Science are related to Statistics and Computer theories, possibly turning Data Science part of the formal sciences.

The scientific method can be applied in Data Science, even partially, contributing to Data Science activities in both domains, scientific and business. In the domain of scientific projects, which are generally guided by the scientific method, less effort is applied in the usage of the method; however, in the business-oriented domain projects, which is more accelerated and competitive, it can sometimes neglect the method and, consequently, the data-based results.

Another relevant consideration regarding the scientific method and Data Science can be observed regarding the use of scientific method that can sometimes be neglected in both domains, due to the avalanche of existing data [10], [13], that processed by powerful algorithms can present satisfactory predictive results, disadvantaging the effective use of the scientific method as part of data science activities.

Applied Data Science serves scientific and business projects, and it also collaborates with a particularity of Data Science activities in both types of projects. Given that the projects in the scientific domain have a greater connection with science and the scientific method, and business projects are carried out in the competitive business world, these differences in both applied Data Science areas can generate different structures of Data Science in the future.

Questions regarding Data Science are still proliferating, and the answers, sometimes, are not conclusive, possibly due to the lack of formalization and foundations regarding Data Science as a new formal knowledge area. The answers to the questions are provided through different statements and biases, depending on the group, community or professionals that provide such answers. Statisticians, Computer Scientists, Business Executives and Data Scientists have different concepts and insights about Data Science. It is even possible to verify a relevant conceptual variation on the term Data Science by itself.

It is also necessary to comment on the characteristics of the professional in the field of Data Science, who has varied profiles as verified in the EDSF [16]. Education and development of these professionals also incurs variations, with no formative standard for Data Science Professional. Although it is possible to observe the effort of the EDISON Project [29] in order to present a standard framework for Data Science profiles and knowledge. Especially as Data Science is at an early stage, graduate programs are more common for Data Science education, with fewer options for undergraduate programs around the world, up to the moment [24]. MOOCs and short-term programs have also been presented as educational options for Data Science professionals.

To conclude, studies, reflections and discussions about formal or applied Data Science are necessary and urgent, but there is a feeling that its progress is slow, especially in the formal area; but, on the other hand, applied Data Science, in the scientific and business domains, is fully working.